\documentclass[aps,prc,floats,floatfix,preprint,superscriptaddress,nofootinbib]{revtex4}  
\def\be{\begin{eqnarray}}    
\def\ee{\end{eqnarray}} 
\usepackage{graphicx} 
 
\begin{document} 
 
\title{Dielectric response of laser-excited silicon} 

\author{S.A. Sato}
\affiliation{Graduate School of Pure and Applied Sciences, University of Tsukuba, Tsukuba 305-8571, Japan}

\author{K. Yabana}
\affiliation{Center for Computational Sciences, University of Tsukuba, Tsukuba 305-8577, Japan}
\affiliation{Graduate School of Pure and Applied Sciences, University of Tsukuba, Tsukuba 305-8571, Japan}

\author{Y. Shinohara}
\affiliation{Graduate School of Pure and Applied Sciences, University of Tsukuba, Tsukuba 305-8571, Japan}

\author{T. Otobe}
\affiliation{Advanced Photon Research Center, JAEA, Kizugawa, Kyoto 619-0615, Japan}

\author{G.F. Bertsch}
\affiliation{Department of Physics and Institute for Nuclear Theory, University of Washington, Seattle 98195, U.S.A.}

\begin{abstract}  
We calculate the dielectric response of crystalline silicon following 
irradiation by a high-intensity laser pulse, modeling the dynamics by 
time-dependent density functional theory (TDDFT).  
The pump-probe measurements are numerically simulated by solving
the time-dependent Kohn-Sham equation with the pump and probe fields
included as external fields.
As expected, the excited silicon shows features of  a particle-hole 
plasma in its 
response.  We compare the calculated response with a thermal model 
and with a simple Drude model.  The thermal model requires only  
a static DFT calculation  to prepare electronically excited
matter and agrees rather well with the TDDFT for the 
same particle-hole density.  The Drude model with two fitted
parameters (electron effective mass and collision time) also  
shows fair agreement at the lower excitation energies; the fitted
effective masses are consistent with carrier-band
dispersions.  
The extracted Drude lifetimes range from 6 fs at weak pumping fields to
much lower values at high fields.  However, we find that the Drude
model does not give a good fit to the imaginary dielectric function
at the highest fields.  Comparing the thermal model with the Drude,
we find that the extracted lifetimes are in the same range, 1-13 fs
depending on the temperature.  These short Drude lifetimes show that
strong damping is possible in the TDDFT, despite the absence of
electron scattering.
One significant difference between the TDDFT response and the other 
models is that the response to the probe pulse depends on the polarization 
of the pump pulse.  We also find that the imaginary part of the dielectric  
function can be negative,  particularly for the parallel polarization  
of pump and probe fields.  
\end{abstract} 
\maketitle 
\section{introduction} 
 
The interaction of high-intensity and ultra-short electromagnetic  
fields with condensed matter is an important subject from both 
fundamental and technological points of view \cite{paris,ps99,bk00,cm07}.  
To investigate dynamics of electrons and phonons in real time,  
the pump-probe experimental technique has been extensively employed. 
One example of its use is creating coherent phonons and measuring their
properties \cite{me97}.
The vibration is detected by measuring the change of reflectance  
of the probe pulse.  However, this requires a good understanding 
of the dielectric  
properties of the surface excited by the pump pulse. 
Another example is the energy deposited by strong laser pulses close to the damage  
threshold.  They  produce high-density electron-hole pairs at the  
surface of dielectrics, causing strong reflection 
for the probe pulse \cite{so00}.  
It is even now possible to measure the population of high-density 
electron-hole pairs in the time resolution less than a 
femtosecond \cite{sb13,sp13,mi11}. 
However, the existing theory
describing these effects is largely phenomenological.
The dielectric properties of laser-excited material are  
often modeled with the Drude model \cite{re04,me10,re10,ka00},  
assuming that excited electrons behave like free carriers. 
 
We have been developing a first-principles theoretical approach  
to describe electron dynamics in crystalline solids induced by  
the intense and ultrashort laser pulses.  
We rely upon the time-dependent density functional theory 
(TDDFT) \cite{rg84},  
solving the time-dependent Kohn-Sham (TDKS) equation in real time  
and real space \cite{yb96}. We have applied our approach to the optical  
breakdown \cite{ot08,ot09}, coherent phonon generation \cite{sh10,sh12}, 
high harmonic generation \cite{ot12}, and coupled dynamics of electrons and  
electromagnetic fields in a multiscale description \cite{ya12}. 
 
In the present paper, we apply the TDDFT to dielectric properties
of a medium excited by short, intense laser pulses.  The method is
to solve the TDKS equation in the medium in the presence of an
external electromagnetic field having both pump and probe pulses.
Thus, we simulate the pump-probe experiments numerically.
The theory describes the electron dynamics fully  
quantum mechanically but assuming that the electrons only interact  
via a time-dependent mean field. Thus, the theory is only expected  
to be justified before the times when electron-electron collisions  
have substantially affected the electronic structure. 
A separate issue is the creation of phonons. For the excitation 
energies we consider here, the electron-electron collision time sets 
a more stringent limit than the phonon interactions. 
 
To interpret the results, we compare with several more simplified 
models.  One model that is often used is based on a Drude response
of the excited quasiparticles, which embedded in a dielectric 
medium \cite{me10,re10}.  
This requires a number of parameters to be fitted.  Another model
ignores the dynamics that created the excited electronic states,
replacing it by a thermal ensemble of electrons \cite{rc06}.  One can carry
out the TDDFT calculation of the linear response of the thermal
system and compare it directly with the response to the pump-excited
system.  We will find that many features of response can be
understood even at a quantitative level with the simpler treatments.
However, there are also  features that only appear in the
full pump-probe simulation.

 The construction of the paper is as follows.
In Sec. \ref{sec:npp}, we describe a method and results of
numerical pump-probe simulation to extract dielectric properties
of excited silicon. In Sec. \ref{sec:thermal}, we present
results of a thermal model and compare them with the numerical
pump-probe results. Our findings are summarized in Sec. \ref{sec:summary}.

\section{Numerical Pump-probe experiments} 
\label{sec:npp}
In this section we carry out what we call numerical pump-probe  
experiments to study the dielectric properties of the highly  
excited material. 
We examine the electronic response in a unit cell of a crystalline
solid irradiated by the pump and probe  
laser pulses. Since the wavelength of the laser pulses is much  
longer than a typical length scale of electron dynamics,  we treat
the laser electric field as a spatially uniform field.  
The current induced by the probe pulse will be used to  
investigate the dielectric properties of excited matter.  
 
\subsection{Calculation of electron dynamics} 
 
Our calculation method has been described in detail 
elsewhere \cite{be00,ya06,be06,sh10,ya12},
so we only provide here the details germane to our study here. 
The electrons dynamics is calculated using the TDKS equation,
\be 
\label{TDKS} 
i\hbar \frac{\partial}{\partial t}\psi_i(\vec{r},t)= 
h_{KS}(t)\psi_i(\vec{r},t), 
\ee 
where $h_{KS}(t)$ is a time-dependent Kohn-Sham Hamiltonian.
It differs from the ordinary TDKS Hamiltonian by inclusion of
the vector potential $\vec A$ in the kinetic term, $p^2/2m \rightarrow
(\vec p+e \vec A/c)^2/2m$.  There is also a coupling to $\vec A$ in the
pseudopotential for the ions; see Ref. \cite{be00,ya06} for details.
The electron-electron interaction in the TDKS
Hamiltonian is modeled in a simple a adiabatic local density 
approximation \cite{lda}.
We calculate dynamics of valence electrons only, treating the 
electron-ion interaction by the norm-conserving pseudopotential 
\cite{tm91,kb82}.
  
The current flowing within the unit cell is given by 
\be 
\label{current} 
\vec{J}(t)=\sum_i \frac{1}{\Omega} \Re \left[ 
\int_{\Omega}d\vec{r} \psi_i^*(\vec{r},t) 
\vec{j}(t) \psi_i(\vec{r},t) \right ], 
\ee 
where $\Omega$ is the volume of the unit cell and the 
current operator $\vec{j}(t)$ is defined by 
\be 
\vec{j}(t)=-\frac{e}{m}\frac{1}{i\hbar}\left[ \vec{r}, 
h_{KS}(t) \right]. 
\ee 
 
The relation of the vector potential 
 $A(t)$ in the unit cell
to the external electromagnetic field exciting the
system depends on a number of factors including possible macroscopic
polarization fields.  In the present analysis, we assume a transverse geometry  
as discussed in Ref. \cite{ya12}.  The sample  
is treated as infinite in the direction of the polarization vector so 
there appears no polarization field inside the solid.  
Of course the field is also affected by the absorption and the reflection
from the surface region, but we don't attempt here to express the results
in terms of the incident laser intensity.
We take the following form for the vector potential  
of the pump pulse in the medium  
$A_{P}(t)$, 
\begin{eqnarray} 
\label{pump-laser} 
A_{P}(t) &=& \nonumber \left\{ \begin{array}{ll} 
    -c\frac{E_0}{\omega_P} \cos{(\omega_P t)}\sin^2(t/{\tau_L})  
& (0<t<{\tau_L}) \\ 
    0 & (otherwise) ,
  \end{array} \right. \nonumber \\ 
\end{eqnarray}
 where $\omega_P$ and $\tau_L$ is the average frequency  and the
time length of the laser pulse, respectively. $E_0$ is the maximum electric 
field strength in the medium.  This is related to the maximum intensity
of the pulse $I$ by  $I_v = c E_0^2/8 \pi $ in the vacuum and
$I_m = \epsilon^{1/2}c E_0^2/8 \pi $ in the medium.  Since
the dielectric function $\epsilon$ is not well-defined in the presence of
a strong electric field, we shall report our results using
the field intensity corresponding to the vacuum relationship.

Our computer code to solve the TDKS equation
uses a three-dimensional grid 
representation to represent orbital wave functions. 
The unit cell for the silicon crystal treated has a length $a=10.26$ a.u. 
contains eight Si atoms. The cubic unit cell is discretized 
into $16^3$ grid points. The four valence electrons of Si atoms 
beyond the closed (1s2s1p) shells are 
treated dynamically. The $k$-space is also discretized into 
$24^3$ grid points. The time evolution is computed using a fourth-order Taylor  
expansion of the operator $\exp(-i h_{KS}(t) \Delta t/\hbar)$ \cite{yb96}.  
We use a time step of $\Delta t=0.08$  
a.u. The number of time step is typically 24,000. 
 
\begin{figure}    
\includegraphics [width = 8cm]{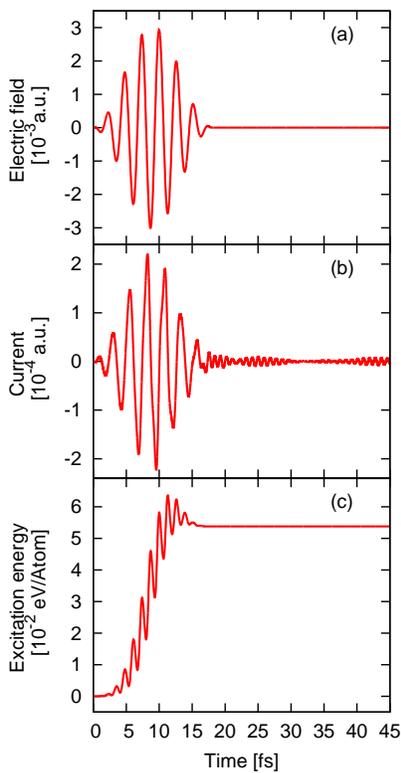}    
\caption{\label{t-dep}  
(a) The time profile of the applied electric field.  
(b) The current induced by the applied electric field. 
(c) The excitation energy per Si atom.  
} 
\end{figure}

In Fig \ref{t-dep}, we show an example of  
the calculated electron dynamics induced by the intense  
pump pulse.  Here the frequency of the pump pulse is set 
to $\hbar\omega_{P}=1.55$ eV and the duration of the 
pump pulse is $\tau_L=18$ fs. These values will be used in all 
calculations of this paper. 
For this figure, the electric field strength corresponds to
an intensity of
$I=3.2 \times 10^{12}$ W/cm$^2$.

Panel a) of the figure shows the time  
profile of the electric field, 
$E_{P}(t)=-\frac{\partial}{c \partial t} A_{P}(t)$. 
Panel b) shows the induced current, calculated using the time-dependent
orbitals in Eq. (\ref{current}). The average frequency
$\hbar\omega=1.55$ eV is smaller than the direct band gap  
energy (2.4 eV in the present calculation), so the initial
current response is nondissipative.  This is seen by the phase 
difference between 
the current and the electric field, which is shifted by $\pi/2$ at  
the beginning  of the field pulse ($t<5$ fs).  
As the intensity of the pulse increases,
the system absorbs energy by the excitation of electron-hole 
pairs.  
As a result, the phase difference decreases.  Note that
the current shows
a weak oscillation at high frequency after the pulse has passed.
Making a Fourier analysis,
we find that it is dominated by frequencies around
3.9 eV/$\hbar$. However, we have no physical explanation of
the oscillation.

Figure \ref{t-dep} (c) shows the  
excitation energy per Si atom. During the field pulse, 
a rapid increase of the electronic excitation  
energy is seen. After the  pulse ends, the excitation  
energy is independent of time, showing that our computational
algorithm conserves the energy of the system. 
 
We next discuss the number density of excited electron-hole pairs and 
the electronic excitation energy when field pulses of different 
intensities irradiate on the silicon crystal. 
To calculate the number density of excited electron-hole pairs, 
we first define eigenstates of the Kohn-Sham Hamiltonian of the  
excited system. We consider a system at a time $t_f$ sufficiently after 
the applied field pulse ends, and denote the Kohn-Sham Hamiltonian 
of Eq. (\ref{TDKS}) at the time $t_f$ as $h_{KS}(t_f)$.  
We introduce the Kohn-Sham orbitals which satisfy the following 
eigenvalue equations. 
\be 
\label{adiabatic} 
\hat{h}_{KS}(t_f) \phi_i^{t_f}=\epsilon_i^{t_f} \phi_i^{t_f} 
\ee 
Note that the Kohn-Sham Hamiltonian $\hat{h}_{KS}(t_f)$ is
different from that of the initial state, due to the change in
electron density in the excited system.
Using eigenfunctions $\phi_i^{t_f}$, we may
define the number density of electron-hole pairs by 
\be 
\label{num-qp} 
n_{eh}=\sum_i \left \{ 
1-\sum_j \left| 
\langle \phi_j^{t_f}|\psi _i(t_f)\rangle 
\right|^2 
\right \}, 
\ee 
where the sum over $i,j$ is taken over occupied orbitals,  
and $|\psi_i(t_f)\rangle$ is the orbital of the TDKS equation at the  
time $t_f$. We use the final-state definition of the orbitals because
it facilitates the comparison to a thermal model that will be discussed
later.

\begin{figure}    
\includegraphics [width = 8cm]{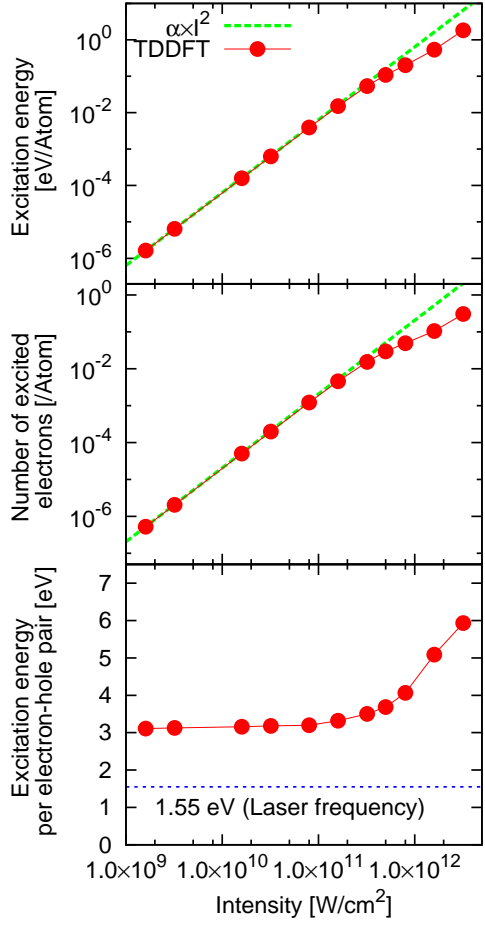}    
\caption{\label{Enex_I}  
Electronic properties of the crystalline silicon in the final state
following the pulsed excitation.  Shown are:
electronic excitation energy per Si atom (top panel);
the number density of electron-hole pairs (middle panel);
the excitation energy per electron-hole 
pair (bottom panel).  Quantities are shown as a function
of the maximum pump intensity determined as $I = c E_0^2/8 \pi$.}
\end{figure}

In Fig. \ref{Enex_I},
we show the number density of electron-hole pairs (top panel) 
and the electronic excitation energy per atom (middle panel). 
The bottom panel shows the electronic excitation energy divided 
by the number of electron-hole pairs. 
As seen from the figure, both excitation energy and the number of  
excited electrons increase with increasing the applied field intensity.  
At low intensity region, they scale with the square of the field 
intensity. This is because two photons are required for electrons 
to be excited across the direct band gap. As seen from the 
bottom panel, the electronic excitation energy per excited 
electron-hole pairs is given by 3.1 eV, which coincides with 
the two-photon energy of the applied field pulse. 
As the field intensity increases above $ 10^{11}$ W/cm$^2$,  
both the number density of electron-hole pairs and the electronic  
excitation energy deviated from the two-photon curve. 
As will be seen later, the dielectric property of excited 
matter also shows a large change from that in the ground state  
at field intensities above this value.

\subsection{Dielectric function from numerical pump-probe calculation} 
 
To extract dielectric properties of excited matter, we compare 
two calculations, one solving the TDKS equation of Eq. (\ref{TDKS})  
with the vector potential containing pump and probe pulses and  
the other containing the pump pulse only.  
We denote the electric field of the pump pulse as $E_{P}(t)$  
and that of the probe pulse as $E_{p}(t)$.  
We denote the current in the calculations containing the pump 
pulse only as $J_{P}(t)$ and that in the calculations containing 
both pump and probe pulses as $J_{Pp}(t)$. 
We define the current induced by the probe pulse as the difference, 
\be 
\label{j-probe} 
J_{p}(t)=J_{Pp}(t)-J_{P}(t). 
\ee 
From the difference of the induced currents, we may extract  
the conductivity and the dielectric function of excited matter. 
 
From the probe current $J_{p}(t)$, we may extract the electric  
conductivity $\sigma(\omega)$ and the dielectric function  
$\epsilon(\omega)$ of excited matter by the following equations:
\be 
\label{conductivity} 
\sigma(\omega)= \frac{\int dt  J_{p}(t)e^{i \omega t}} 
{\int dt E_{p}(t)e^{i \omega t}}  
\ee 
\be 
\label{dielectric} 
\epsilon(\omega)=1+\frac{4\pi i \sigma(\omega)}{\omega}, 
\ee 
In principle, the above-defined conductivity and dielectric function   
depend also on the time delay $\tau_{Pp}$ between the pump and probe pulses. 
We will later show that the dependence  
on delay-time is rather small in the TDDFT calculations. 
 
In practice we employ the vector potential of  
the form of Eq. (\ref{pump-laser}) as the pump pulse. 
As for the probe pulse, we use the same functional form as 
Eq. (\ref{pump-laser}) delayed by an amount $\tau_{Pp}$ from the pump pulse, 
\be 
\label{probe} 
A_{p}(t)&=&-c\frac{e_0}{\omega_{p}} \cos{(\omega_{p}(t-\tau_{Pp}))} 
\nonumber \\ 
&&\times \sin^2((t-\tau_{Pp})/\tau_L) 
\ee 
for $\tau_{Pp}<t<\tau_L+\tau_{Pp}$ and zero otherwise. 
 
\begin{figure}    
\includegraphics [width = 9cm]{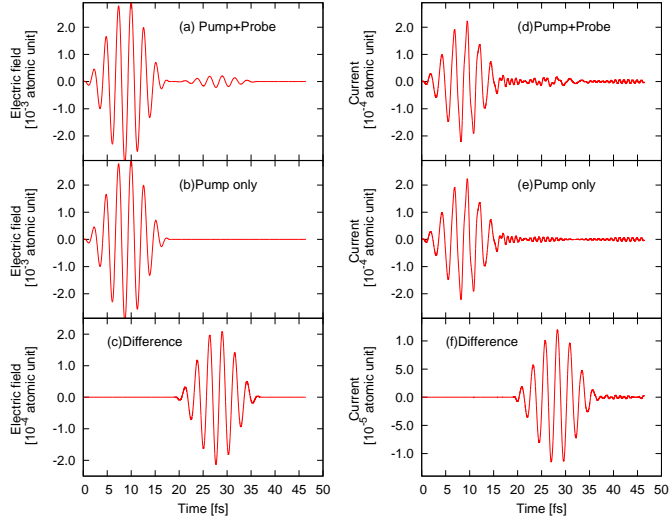}    
\caption{\label{pp_Ej}  
Left panels show the electric field of pump and probe pulses in (a), 
pump pulse in (b), and the probe pulse in (c). Right panels show 
the current induced by the pump plus probe pulse in (d), the current 
by the pump pulse only in (e), and the difference of the currents 
shown in (d) and (e) in (f).} 
\end{figure}

In Fig. \ref{pp_Ej},  
we show typical time profiles of the electric fields and  
the induced currents for a delay time of  $\tau_{Pp}=19$ fs.  
The pump pulse is the same as in Fig. 1, with a maximum intensity
of $3.2 \times 10^{11}$ W/cm$^2$.  
The probe intensity is a factor
of 200 smaller, which we deem to be sufficiently weak to extract
the linear response.
In the left panels of  
Fig. \ref{pp_Ej}, 
we show electric fields of pump and probe pulses, $E_{P}(t)+E_{p}(t)$,  
in (a), pump pulse, $E_{P}(t)$, in (b), and probe pulse, 
$E_{p}(t)$, in (c), as functions of the time. 
The right panels show currents induced by the pump and probe pulses, 
$J_{Pp}(t)$, in (d), by the pump pulse only, $J_{P}(t)$, in (e),  
and the difference of the currents, $J_p(t)$ of Eq. (\ref{j-probe}),  
in (f). 
 
The next step is to calculate the dielectric function from the  
probe current using Eqs. (\ref{conductivity}) and (\ref{dielectric}).  
The pump-probe calculation using the probe pulse of  
Eq. (\ref{probe}) and the probe current of Eq. (\ref{j-probe})  
gives us dielectric properties around the average frequency  
$\hbar\omega_{p}$. To investigate dielectric properties  
for a wide frequency region, we repeat the pump-probe calculations 
for different frequencies of the probe pulses.
 
\begin{figure}    
\includegraphics [width = 8cm]{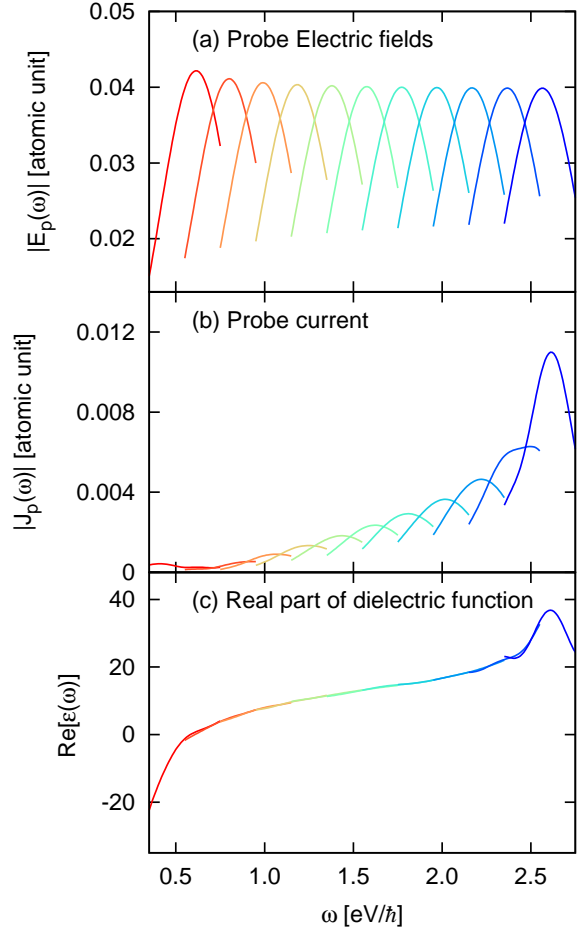}    
\caption{\label{pp_FT}   
The top two panels show the Fourier transformations of the probe  
electric field $E_{p}(\omega)$ and the probe current  
$J_{p}(\omega)$. The bottom panel shows the real part of the  
dielectric function extracted from $E_{p}(\omega)$ and  
$J_{p}(\omega)$ through Eqs. (\ref{conductivity}) and 
(\ref{dielectric}).  The pump pulse has
an intensity $I=3.2 \times 10^{12}$ W/cm$^2$ and an average frequency
$\omega_P=1.55$ eV/$\hbar$. The polarization directions of 
the pump and probe pulses are taken to be parallel.
} 
\end{figure}

In Fig. \ref{pp_FT}, 
we show typical calculations using a number of probe pulses of
differing frequencies.
Panels (a)  
and (b) show the absolute values of the Fourier transforms 
of $E_{p}(t)$ and $J_{p}(t)$, respectively. 
Panel (c) 
shows the real part of the dielectric function which is 
calculated using Eqs. (\ref{conductivity}) and (\ref{dielectric}).  
The curve is composed of a number of curves with different  
colors for each probe frequency.  One can see that the overlap is
very good for the different average probe frequencies, validating
our method to extract the dielectric function.  

\begin{figure}    
\includegraphics [width = 9cm]{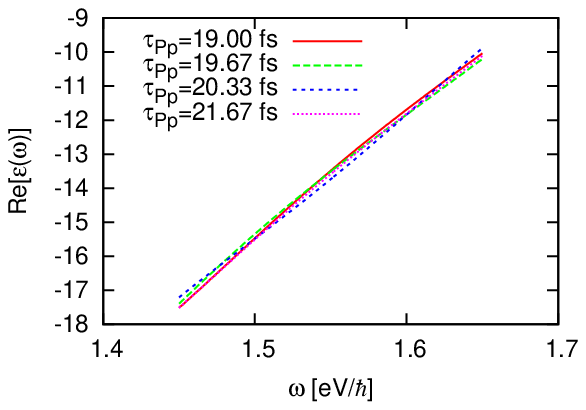}%
\includegraphics [width = 9cm]{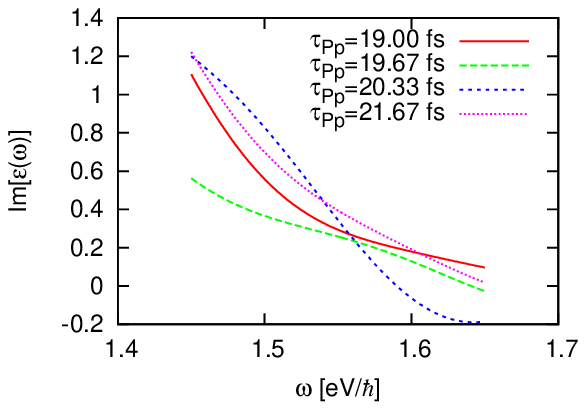}
\caption{\label{delay}  Extracted dielectric functions as a 
function of the delay time $\tau_{Pp}$.  The pump pulse has
an intensity $I=3.2 \times 10^{12}$ W/cm$^2$ and an average frequency
$\omega_P=1.55$ eV/$\hbar$.  The probe frequency is $\omega_p = 1.55$
eV/$\hbar$ and its delay times for the four graphed lines
are $\tau_{Pp} = 19,0, 19.67, 20.33$ and $21.67$ fs.
 The polarization directions of the pump and probe pulses are
taken to be parallel.
} 
\end{figure}

We next ask how sensitive is the extracted dielectric function to
the time delay of the probe pulse?  Since there are no dissipative
processes in the TDDFT  under ALDA, 
the properties of the system should not
change after some initial period when the phases of the excited
orbitals become incoherent.  Figure \ref{delay} shows how the
extracted dielectric function depends on delay time for one of the
cases.  
We have selected delay times over a range that corresponds to a full
cycle of the pump pulse, since that frequency could be imprinted on 
the phases of the particle at later times.  The range of the delay
times is 19.00 fs, 19.67 fs, 20.33fs, 
and 21.67fs. The latter three delay times correspond 
to a quarter, a half, and one period of the pump pulse 
$2\pi/\omega_P$  added to the first time.
One can see that real part is practically independent of 
the delay.  The imaginary part, however, shows variation although
qualitatively the functions are similar.  We found the same behavior
for other cases as well.  Namely, the real part is 
independent of delay, even extending the delay to very large
times.  The imaginary part
is only qualitatively similar for different delay times.
In the sequel, we will analyze all the results using the dielectric
function at $\tau_{Pp}=19 $ fs, and one should remember that the imaginary
part is less well defined than the real part.

\begin{figure}    
\includegraphics [width = 8cm]{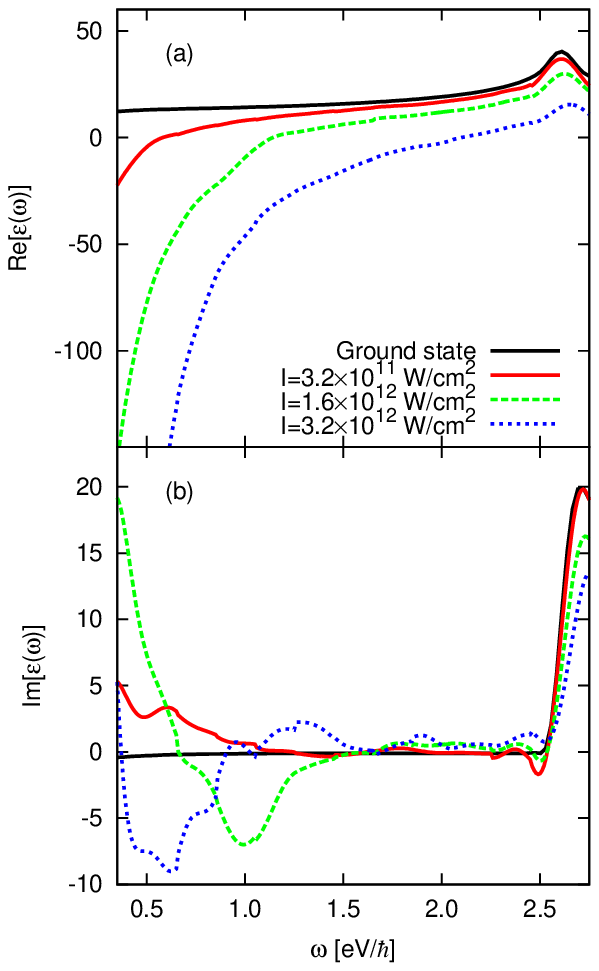}    
\caption{\label{eps_pp}  
Real (a) and imaginary (b) parts of the dielectric functions 
of Si excited by field pulses of three intensities. 
The dielectric functions are deduced using pump-probe calculations.  
The polarization direction of the probe pulse is taken to be 
parallel to that of pump pulse. The dielectric function 
of silicon in the ground state is also shown. 
} 
\end{figure} 
 
We have carried out the pump-probe simulations for several intensities
of the pump pulse.  The results for the dielectric functions are
shown in Fig \ref{eps_pp}. The real and the imaginary  
parts are presented in panels (a) and (b), respectively. 
 
The distinguishing feature in the response is the negative
divergence at small frequencies.   This arises from the quasiparticles 
in the excited system,  
as we will see more quantitatively later.  
The imaginary part of the response is not quite as simple to analyze.  
The plasmon in a free electron gas at zero temperature is undamped  
and the imaginary part of the dielectric function vanishes at low frequency.  
Figure \ref{eps_pp} shows, however, that  
${\rm Im}[\epsilon (\omega )]$ becomes large at low frequency.  
The quasiparticle response is thus far from that of a free electron gas.
 
\begin{figure}    
\includegraphics [width = 10cm]{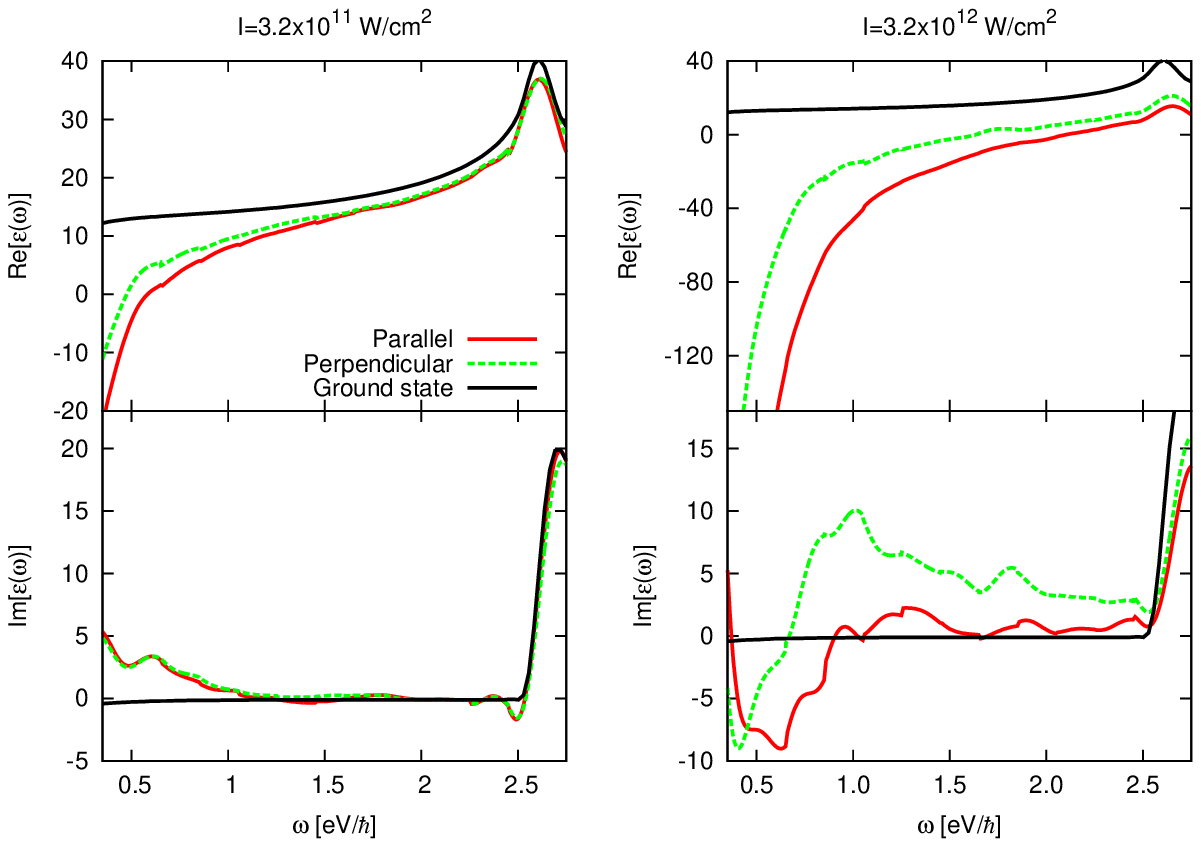}    
\caption{\label{eps_para_prep}  
Comparison of dielectric functions of excited silicon probed with either 
parallel or perpendicular direction to the pump fields. 
} 
\end{figure}

An interesting feature of the TDDFT response is that the dielectric  
tensor is not isotropic in the excited crystal, even though the crystal  
symmetry is cubic. This may be seen in Fig. \ref{eps_para_prep},  
comparing the dielectric functions for the probe polarization either  
parallel or perpendicular to the pump polarization.  
 
The real part of the dielectric function shows the low-frequency  
plasmon response more strongly for the parallel component.  
One may notice that ${\rm Im}[\epsilon (\omega )]$ is negative  
at some frequencies. This might indicate a population  
inversion that could sustain a growth of intensity at those frequencies.  
However, one should carry the full calculation of the pulse propagating  
in space as well as time to assert that the excited medium can amply  
the pulses.

\subsection{Comparison with free-carrier models} 
 
Dielectric properties of solids excited by intense and  
ultrashort laser pulses are often modeled employing a simplified  
dielectric function, adding a Drude-like component to the  
dielectric function in the ground state \cite{me10,re10}. In this subsection,  
we will examine how well the dielectric function of excited  
matter in the TDDFT calculation may be described by a
simplified dielectric model. 
 
First we consider an embedded Drude model, the dielectric 
function given as a sum of the ground state response and the  
Drude response of free carriers  
\be 
\label{eb-drude} 
\epsilon_{ED}(\omega)=\epsilon_0
(\omega)-4\pi i \frac{e^2n_{eh}}{m^*\omega(\omega+i/\tau)}.
\ee 
Here $\epsilon_0(\omega)$ is the dielectric function  
in the ground state, $n_{eh}$ is the electron-hole density,  
$m^*$ is the reduced mass of electron-holes, and $\tau$ is  
the Drude damping time.  
For the dielectric function in the ground state, 
$\epsilon_0(\omega)$, we will use the values obtain from the TDDFT calculation. 
The number density of electron-hole pairs, $n_{eh}$, is extracted
from the calculation using Eq. (\ref{num-qp}).  
We treat $m^*$ and $\tau$ as parameters, fitting to
the calculated $\epsilon(\omega)$.
 
Sokolowski-Tinten and  von der Lind proposed a more complicated  
model for the dielectric function excited by strong laser fields \cite{so00},
which we shall call the SL model.  
They consider three physical  
effects for the dielectric response of laser-excited semiconductor:  
(i) state and band filling, (ii) renormalization of the band structure,  
and (iii) the free-carrier response. The SL dielectric function is 
parameterized as 
\be 
\epsilon_{SL}(\omega) &=& 1+\left [ 
\epsilon_0 (\omega+\Delta E_{gap})-1 \right ] 
\frac{n_0-n_{eh}}{n_0} \nonumber \\ 
&&-4\pi i \frac{e^2n_{eh}}{m^*\omega(\omega+i/\tau)}, 
\ee 
where $\Delta E_{gap}$ is the change of the band gap by the laser  
irradiation and $n_0$ is the density of electrons which contribute  
to the dielectric response.  
For $\Delta E_{gap}$, we use a change of single particle energies, 
$\epsilon_i^{t_f}$ of Eq. (\ref{adiabatic}), after the laser pulse ended. 
We treat the active number of valence electrons, $n_0$, as a fitting parameter. 
 
\begin{figure}    
\includegraphics [width = 10cm]{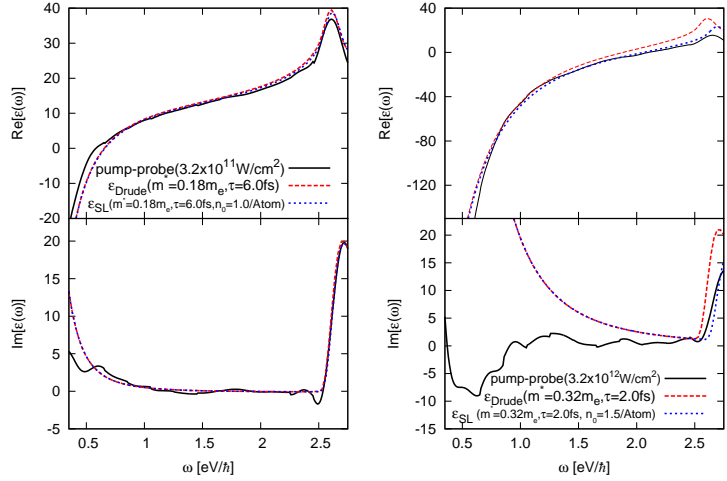} 
\caption{\label{eps_drude}  
Dielectric function of silicon excited by two field pulses, 
$3.2 \times 10^{11} {\rm W/cm^2}$ (left panels) and 
$3.2 \times 10^{12} {\rm W/cm^2}$ (right panels), 
are fitted by the embedded Drude model and by the  
SL model.} 
\end{figure}    
 
Figure \ref{eps_drude} shows the fits obtained in the embedded Drude
model and the SL model for two intensities of the pump field.
 The polarization directions of the pump and probe pulses are
taken to be parallel.
At the lower intensity, the dielectric functions 
are well fitted by both models.  At the higher intensity,
the real part is well described by both models at lower frequencies, but
the SL model fits better above the direct band gap.  However,
neither model does well for the imaginary part in the high
intensity case.
 
In the fitting procedure, we found the effective mass is  
sensitive to the real part of the dielectric function and  
can be determined without ambiguity. 
The effective mass for the pump pulse of 
$3.2 \times 10^{11}$ W/cm$^2$ is given by $m^* = 0.18$,  
while the effective mass for the pump pulse of  
$3.2 \times 10^{12}$ W/cm$^2$ is given by $m^*=0.32$. 
The collision time for the weaker pump pulse case of  
$3.2 \times 10^{11}$ W/cm$^2$ is determined to be about 6 fs.  
However, the Drude damping  
time $\tau$ cannot be  
determined well for high intensity case, because 
the frequency dependence of the 
imaginary-part dielectric function is very different  
from the Drude behavior.  

The effective mass and its change with excitation energy may be  
understood from the band structure. A weak pump  
pulse excites electrons at specific $k$-points by  
two-photon absorption, while a strong pump pulse excites  
electrons at various $k$-points by tunnel and multi-photon  
excitations. The effective mass of electrons 
depends very much on the their positions in the bands; only
the lowest bands have the very small effective masses.
 
\section{Thermal model} 
\label{sec:thermal}
 
The numerical pump-probe experiments reported in the 
previous section are applicable to 
the excited matter immediately after the 
pump irradiation, perhaps for a time period of a few tens of femtoseconds.
In this section,  
we will investigate dielectric properties of  
thermally excited matter, which should be more appropriate at later
times.  We assume that the electronic states are described by a thermal
ensemble, but the ions have not yet had time to respond.  
 
\subsection{Dielectric function at finite temperature} 
 
We describe the thermally excited matter by static density 
functional theory at finite electron temperature. 
In the calculations, the electrons population is described by  
the Fermi-Dirac distribution of a given temperature, while 
atomic positions are frozen at the ground-state positions.
The electron density at temperature $T$, $\rho^T(\vec{r})$, 
is given by  
\be 
\rho^T(\vec{r})=\sum_i n_i^T |\phi_i(\vec{r})|^2 , 
\ee 
where $n_i^T$ is the temperature-dependent occupation number  
of Fermi-Dirac distribution, 
\be 
n_i^T=\frac{1}{1+{\rm e}^{(\epsilon_i-\mu)/k_B T}}, 
\ee 
Here $\epsilon_i$ is the single particle energy, $\mu$ is  
the chemical potential, and $k_BT$ is the temperature in 
energy units. We note that all the quantities related
to the orbitals, $\phi_i$, $\epsilon_i$, and $\mu$ depend
on the temperature $T$ 
due to the self-consistency requirement.
 
\begin{figure}    
\includegraphics [width = 8cm]{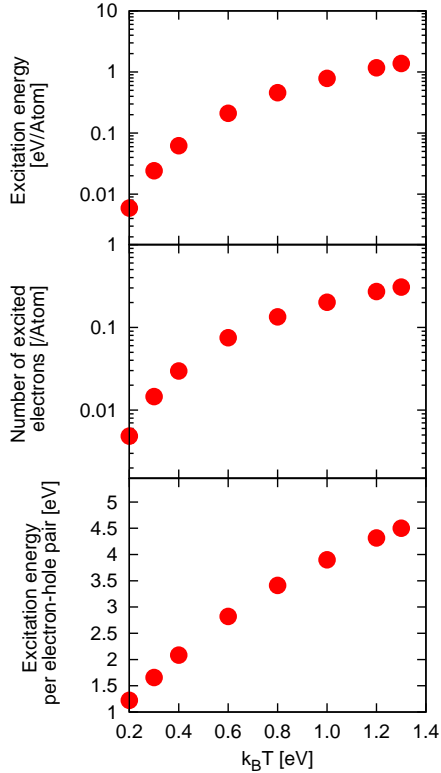}    
\caption{\label{Enex_thermal}  
Basic electron properties of the thermal model of crystalline silicon.
The top panel shows the excitation energy per Si atom.  
The middle panel shows the number of excited electron per Si atom.  
The bottom panel shows the excitation energy per excited electron.  
} 
\end{figure}

In Fig. \ref{Enex_thermal}, 
we show calculated results at several electron temperatures. 
The top panel (a) shows the excitation energy per atom, 
the middle panel (b) shows the number of excited electrons, 
and the bottom panel (c) shows the excitation energy per excited 
electron. As seen from the figure, both excitation energy and the 
number of excited electrons monotonically increase as the 
electron temperature increases. At very low temperature, 
we find the excitation energy per excited electrons is rather 
small, 1.2 eV for $k_B T=0.2$ eV. This value should approach 
to the energy of the indirect gap, 0.52 eV in our calculation, 
in the low temperature limit. 
 
\begin{figure}    
\includegraphics [width = 8cm]{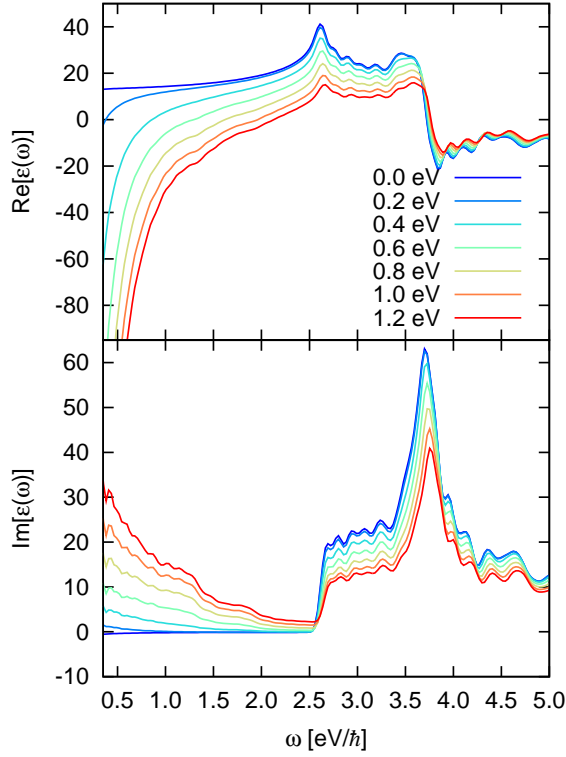}    
\caption{\label{eps_thermal} 
The dielectric functions of the finite temperature model  
at several temperatures. Top panel shows the real part of  
the dielectric function, and the bottom panel shows the  
imaginary part.  
} 
\end{figure}

We investigate dielectric properties of crystalline silicon at finite temperature  
by the real-time method, applying a distorting vector potential  
of step function in time \cite{be00}. 
In Fig. \ref{eps_thermal}, we show dielectric functions of  
silicon at several electron temperatures.  
As seen from the real-part of the dielectric function,  
all responses at finite temperatures show a Drude-like like  
behavior at low frequencies. This behavior is more or less similar  
to those in our numerical pump-probe calculations shown  
in Fig. \ref{eps_pp}. 
The low energy component of the imaginary part shows absorptive 
contributions at low frequencies, increasing monotonically 
as the temperature increases.

\begin{figure}    
\includegraphics [width = 8cm]{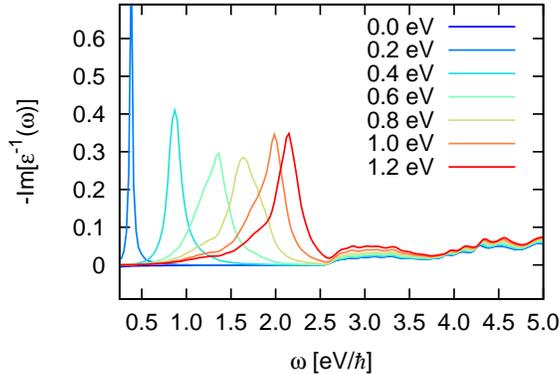}    
\caption{\label{ineps_thermal}  
The imaginary part of the inverse dielectric function for 
various electronic temperatures. 
} 
\end{figure}

A good way to display a plasmon contribution to the response
is to plot the imaginary part of the inverse dielectric function, ${\rm Im} 
\epsilon^{-1}$.  This is shown in Fig. \ref{ineps_thermal} for
several temperatures up to $k_B T = 1.2$ eV.  At the lowest
temperature one sees a very sharp plasmon peak, located at
an energy of $\sim 0.4$ eV.  The plasmon excitation energy 
increases with temperature, due to the increased density of 
electron-hole pairs. 
We note that the width of the plasmon also increases with  
temperature, up to about $k_B T \approx 0.6 $ eV.  Beyond that,
the width does not change very much, up to the maximum temperature
considered.
 
\begin{figure}    
\includegraphics [width = 8cm]{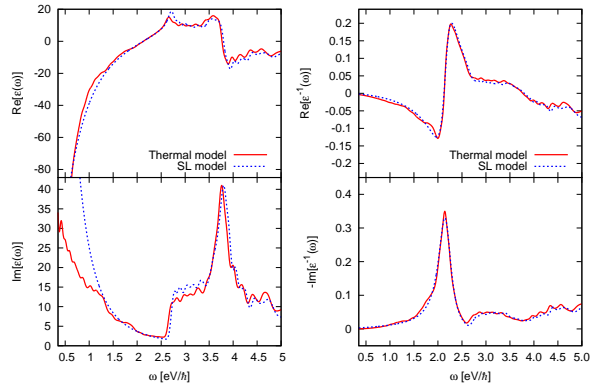}    
\caption{\label{eps_thermal_drude}  
Comparison of the thermal model and a fit with the 
SL model. The electronic temperature in the thermal 
model is $k_B T = 1.2$ eV. 
} 
\end{figure}

The thermal dielectric function presented here 
can be well fitted by the SL model. For the basic parameters of 
the SL model, we employ the calculated dielectric function  
at zero temperature for $\epsilon_0$, the calculated  
electron-hole pair density for $n_{eh}$, and the calculated  
shift of the gap energy for $\Delta E_{gap}$. Other three  
parameters, $m^*$, $\tau$, and $n_0$ are treated as fitting  
parameters. The fit is carried out by minimizing the mean square
error as given by
\be 
I_{error}=\int_{\omega_i}^{\omega_f}d\omega \left| 
\epsilon_{T}^{-1}(\omega)-\epsilon_{SL}^{-1}(\omega) 
\right|^2, 
\ee 
in the interval $\hbar \omega_i= 0.35$ eV and $\hbar \omega_f=5.0$  
eV. The $\epsilon_{T}(\omega)$ is the dielectric  
function in the thermal model.  The quality of the fit is
shown in Fig. \ref{eps_thermal_drude}  for a temperature
of $k_B T= 1.2$ eV in the thermal model.  The fit is very good
except  for the ${\rm Im} \epsilon$ at the lowest
frequencies. In particular, the  
plasmon feature in the inverse dielectric function is very
well reproduced.
 
\begin{figure}    
\includegraphics [width = 8cm]{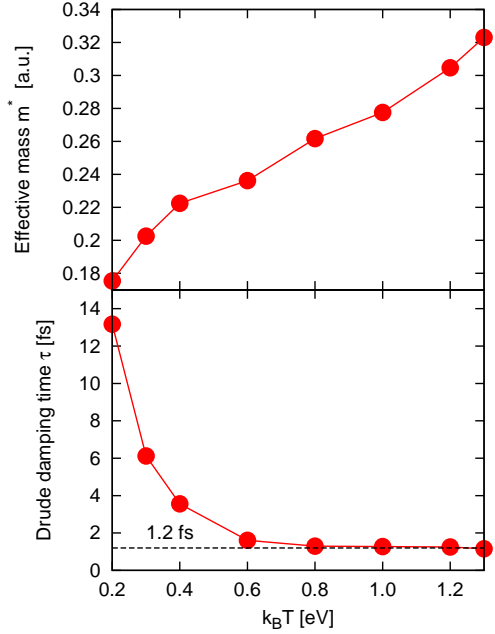} 
\caption{\label{mtau_thermal} 
Parameters of the SL model determined by the fitting procedure 
to the thermal model. 
Top panel shows the effective mass $m^*$ and the bottom panel  
shows the Drude damping time $\tau$. 
} 
\end{figure}

In Fig. \ref{mtau_thermal}, we show the fitted effective mass 
$m^*$ and the collision time $\tau$ as functions of the temperature 
in the thermal model. 
The top panel shows that the effective mass $m^*$ gets heavier  
as the temperature increases. This feature was also found for the
dielectric function extracted from the
numerical pump-probe  
experiments, see Fig. \ref{eps_drude}. 
As we discussed in Sec. \ref{sec:npp}, the change of effective mass may 
be understood by the change of the distribution of the  
electron-hole quasiparticles in $k$-space. 
 
The bottom panel of Fig. \ref{mtau_thermal} shows that 
the damping time $\tau$ becomes very small as the electron temperature  
increases. The value of $\tau$ reaches a saturated value of 1.2 fs 
at $k_B T \approx 0.8$ eV.  At first sight this is puzzling, because
there are no collision 
effects in either the TDKS equation or in the thermal model. In spite of 
this fact, we obtain  
a plasmon feature with large damping, corresponding to collision
times as short as 1.2 fs in the thermal model. We believe that 
the damping 
arises from the elastic scattering of high-energy quasiparticles from 
ionic core  potentials, but we have no quantitative understanding of its 
magnitude or dependence on the quasiparticle distribution.
We note that TDDFT treatment of linear response  describes
the dielectric function of metals fairly well, including the width
of plasmon seen in the inverse dielectric function \cite{be00}.

\subsection{Comparison with numerical pump-probe experiments} 
 
The difference between the numerical pump-probe calculations  
presented in Sec. \ref{sec:npp} and the thermal model comes  
entirely from the different electron-hole distributions  
in the excited system to be probed. In this subsection, we compare 
their predicted dielectric functions.
 
\begin{figure}    
\includegraphics [width = 10cm]{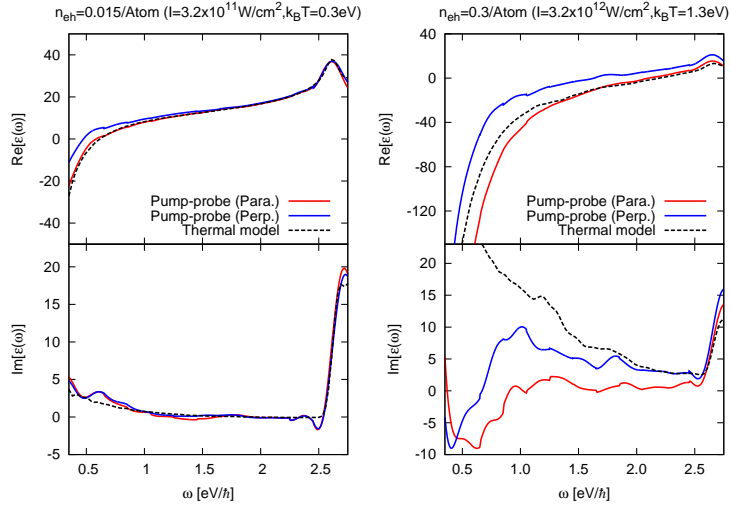}    
\caption{\label{eps_pp_thermal}  
Comparisons of the dielectric function of the numerical pump-probe 
calculation and the thermal model.  Left-hand panels:  $n_{eh} = 0.015$
/Atom; right-hand panels:  $n_{eh} = 0.3$ /Atom.
} 
\end{figure}

We first need to assume a correspondence between  
the excited systems that we wish to compare.  Two possibilities
come to mind, namely consider systems of equal excitation energy or 
of equal densities of particle-hole excitations.  In general, the
laser-excited system will have a higher excitation energy for the
same number of particle-hole pairs.  Since the plasmon characteristics
are closely tied to the number of quasiparticles, we shall use that
measure to make the comparison.  One comparison will be with 
$n_{ph} = 0.015$ /Atom; this is obtained by a pump pulse of 
$3.2 \times 10^{11}$ W/cm$^2$ or by a thermal system with electron
temperature $k_B T= 0.3$ eV.  Another comparison will be with
$n_{ph} = 0.3$ /Atom, requiring  a pump pulse of 
$3.2 \times 10^{12}$ W/cm$^2$ or a temperature of $k_B T= 1.3$ eV.
The two dielectric functions are shown in Fig.  \ref{eps_pp_thermal}. 
 
As seen from the figure, the dielectric function of silicon excited  
by the low-intensity pump-pulse ($I=3.2 \times 10^{11}$ W/cm$^2$) 
is close to the thermal model.  
At the stronger pump intensity of $I=3.2 \times 10^{12}$ W/cm$^2$, 
the real part of the dielectric function in the thermal 
model lies between the perpendicular and parallel dielectric functions 
of the numerically pumped system.  In fact,
the finite temperature function is close to the parallel case at the
higher frequencies. 
On the other hand, 
the imaginary part of the dielectric function agrees with the perpendicular
case at high frequencies, but is much larger than either polarizations
at low frequency.
We thus conclude that, making correspondence between the numerical 
pump-probe calculations and the thermal electron model using the 
number density of electron-hole pairs, the dielectric functions show 
a reasonable correspondence, especially when the excitations are 
not very violent. The difference between two calculations 
comes from the different $k$-space distributions of electron-hole 
pairs. It seems that the difference is more evident for the imaginary 
part.

\section{SUMMARY} 
\label{sec:summary}
 
To investigate a change of dielectric properties of bulk silicon 
immediately after ultrashort laser pulses, we made  
numerical pump-probe experiments solving the time-dependent  
Kohn-Sham equation in real time including electric fields  
of both pump and probe pulses. The simulation makes it  
possible to investigate dielectric properties of excited  
matter before any dissipation or dephasing effects start to 
become significant.
 
For a comparison, we have also constructed a thermal model
by solving the static Kohn-Sham equation with finite-temperature
Fermi function occupation factors.  Its dielectric response was then computed
by applying the linear response theory using the usual  real-time
method.  We found that the thermal model works very well at
lower excitation energies, but becomes unreliable at the higher
excitation energy where one finds differences in the parallel
and perpendicular dielectric functions.  
 
An even more simple model can be constructed using ingredients
of the Drude model of free-electron dynamics.  
In general, the real part of the dielectric function  
was found to be well described by a Drude-like contribution
of the excited quasiparticles  
embedded in the dielectric medium corresponding to the ground state.  
As for the imaginary part, the dielectric function in the  
thermal model is reasonably described by the embedded Drude model.  
The dielectric function 
in the numerical pump-probe experiments shows rather 
different behavior, even negative values for the imaginary 
part. The difference comes from the nonequilibrium 
distributions of electrons and holes in $k$-space in the 
numerical pump-probe experiments. 
 
In the embedded Drude model, there are three parameters determining
the quasiparticle plasmon contribution, namely the density of 
quasiparticles, their effective mass $m^*$, and the collision 
time $\tau$.
The density of quasiparticles is known from the TDDFT or thermal 
calculation, but the other quantities are fit.
From the real part of the dielectric function, we find  
increase of the effective mass as the pump field intensity  
increases as expected from the band structure. 
We find the calculated dielectric functions  
show substantial imaginary part. In the thermal electron 
model, the collision time of as short as 1.2 fs gives  
reasonable fit. This short value for the collision time  
is unexpected, since there are no explicit collision terms
in the time-dependent  
Kohn-Sham equation that we solve.
We believe the short collision time comes from the  
elastic scattering of electrons from atoms rather than 
from electron-electron or electron-phonon interactions, 
but we lack a simple model to exhibit this aspect of 
the response.

\section*{Acknowledgments}

This work is supported by the Grants-in-Aid for Scientific Research No.
23340113, No. 23104503, No. 21340073, and No. 21740303. The numerical
calculations were performed on the supercomputer at the Institute of Solid
State Physics, University of Tokyo, and T2K-Tsukuba at the Center for
Computational Sciences, University of Tsukuba. G.F.B. acknowledges support
by the National Science Foundation under Grant No. PHY-0835543 and by the US
Department of Energy under Grant No. DE-FG02-00ER41132.

\end{document}